\documentstyle[psfig]{article}
\title{\bf Gauge fixed domain wall fermions on lattice at small
Yukawa coupling}
\author{S. Basak$^{1,2}$ and Asit K. De$^1$ \\
$^1$Theory Group, Saha Institute of Nuclear Physics, \\
1/AF Salt Lake, Calcutta 700064, India. \\
$^2$Dept. of Physics, NND College, Calcutta 700 092, India.}
\date{}

\begin{document}
\maketitle

\begin{abstract}
Gauge fixed domain wall fermions are investigated in the reduced
model at small Yukawa couplings.  We present chiral propagators
at the waveguide boundaries using quenched numerical simulations
and analytic methods. There is no evidence of mirror chiral modes
at the waveguide boundaries.
\end{abstract}

\section*{Introduction} \label{intro}
Recently it has been shown \cite{bock1,bd} that nonperturbative
gauge fixing \cite{golter1} can be applied successfully to decouple
the longitudinal gauge degrees of freedom from Abelian lattice chiral
gauge theories in the limit of zero gauge coupling. After one gauge
transforms a gauge-noninvariant lattice chiral gauge theory proposal
like the Smit-Swift model or the domain wall waveguide model, one
picks up the longitudinal gauge degrees of freedom (radially frozen
scalars) explicitly in the action. The job of gauge fixing is to
find a phase transition where the gauge symmetry would be restored
with the scalars decoupled. 

The gauge fixing approach deals with anomaly-free chiral gauge
theories. It is manifestly local. The gauge fixing action involves
a kinetic term for the longitudinal gauge degrees of freedom and
allows for a renormalizable weak coupling expansion.

The so called {\em reduced} model is defined by taking the limit
of zero gauge coupling, after gauge transforming the
gauge-noninvariant theory. Hence it has coupling between fermions
and the scalars. The spectrum of the reduced model derived from
the gauge fixed theory should be devoid of the scalars and consist
only of free fermions in the appropriate representation of the
chiral group. This is precisely what was achieved with recent
investigations of Smit-Swift \cite{bock1} and domain wall models
\cite{bd} and tuning counterterms posed no practical problems.
In the reduced model there is no constraint on the fermion
representation from anomalies since the anomaly vanishes trivially
in this case, and one can still study the fermion spectra.

In this communication we extend our detailed study \cite{bd} of
the U(1) lattice chiral gauge theory with domain wall fermions
and gauge fixing in the reduced model limit. The reduced model
has a Yukawa coupling $y$ which was put in by hand. Our previous
study dealt with the situation at $y=1$ and showed that the scalar
fields were completely decoupled from the fermions and chiral
modes were obtained only at the domain wall and at the anti-domain
wall as in free domain wall fermions. This makes the gauge fixed
domain wall model suitable for a U(1) chiral gauge theory at $y=1$.
On the other hand, at $y=0$, fermion current considerations and
also numerical simulations show that mirror chiral modes develop
at the so-called waveguide boundaries making it unsuitable for
the construction of a chiral gauge theory. Hence it is interesting
to ask what happens at small nonzero values of Yukawa coupling. 
In general it is desirable to understand the properties of a
model for all possible values of the coupling parameters. Here we
investigate the gauge fixed domain wall waveguide model in the
reduced limit for small nonzero values of the Yukawa coupling.
 
\section*{Gauge fixed Domain Wall Action} \label{gfdwall}
Kaplan's free domain wall fermions \cite{kaplan1} on a
$4+1$-dimensional $L^4L_s$ lattice ($0\le s\le L_s-1$, $s$
labeling the 5-th dimension) with periodic boundary conditions
in the 5-th direction and the domain wall mass taken as
\begin{equation}
m(s) =  \begin{array}{rl}
-m_0, &  0 <s< L_s/2 \\
 0,   &  s = 0, L_s/2 \\
 m_0, &  L_s/2 <s< L_s
\end{array} \label{dwmass}
\end{equation}
possess a lefthanded (LH) chiral mode bound to the domain wall
at $s=0$ and a righthanded (RH) chiral mode bound to the anti-domain
wall at $s=L_s/2$. For $m_0 L_s\gg 1$, these modes have exponentially
small overlap.

A 4-dimensional gauge field which is independent of $s$ is
then coupled to fermions only for a restricted number of $s$-slices
around say, the anti-domain wall \cite{kaplan2,golter2} with a view
to coupling only to the RH mode at the anti-domain wall. The gauge
field is thus confined within a {\em waveguide},
\begin{equation}
WG = (s: s_0 < s \leq s_1) .
\end{equation}
The waveguide boundaries $s_0$ and $s_1$ should be reasonably 
far away from the domain wall and the anti-domain wall.

Obviously, the hopping terms from $s_0$ to $s_0+1$ and that from
$s_1$ to $s_1+1$ would break the local gauge invariance of the
action. This is taken care of by gauge transforming the action and
thereby picking up the longitudinal gauge degrees of freedom or
radially frozen scalar fields $\varphi$ at the waveguide boundary,
leading to the action (lattice constant is taken to be unity throughout):
\begin{eqnarray}
S_{F} & = & \sum_{s \in WG}
\overline{\psi}^s \left( D\!\!\!\!/\,(U) - W(U) + m(s) \right) \psi^s
+ \sum_{s\not\in WG} \overline{\psi}^s \left( \partial\!\!\!/
- w + m(s) \right) \psi^s \nonumber \\
&+& \;\sum_s \overline{\psi}^s \psi^s - \sum_{s\neq s_0,s_1}
\left( \overline{\psi}^s P_L \psi^{s+1} + \overline{\psi}^{s+1} P_R
\psi^s \right) \nonumber \\
&-& \;y \left( \overline{\psi}^{s_0} \varphi^\dagger P_L \psi^{s_0+1} +
\overline{\psi}^{s_0+1} \varphi P_R \psi^{s_0} \right)
- y \left( \overline{\psi}^{s_1} \varphi P_L
\psi^{s_1+1} + \overline{\psi}^{s_1+1} \varphi^\dagger P_R \psi^{s_1}
\right) \label{wgact}
\end{eqnarray}
where we have suppressed all indices other than $s$. $\overline{\psi}$
and $\psi$ are the fermion fields, the projector $P_{L(R)}$ is
$(1\mp \gamma_5)/2$ and $y$ is the Yukawa coupling introduced by
hand at the waveguide boundaries. $D\!\!\!\!/\;(U)$ and $W(U)$ are
respectively the gauge covariant Dirac operator and the Wilson term
(with Wilson $r=1$) in 4 space-time dimensions. $\partial\!\!\!/$
and $w$ are the 4-dimensional free Dirac and Wilson operators.

The gauge fixed pure gauge action for $U(1)$, where the ghosts
are free and decoupled, is:
\begin{equation}
S_B(U) = S_g(U) + S_{gf}(U) + S_{ct}(U) \label{ggact}
\end{equation}
where, $S_g$ is the usual Wilson plaquette action and $S_{ct}$ is a 
counterterm. Previous studies \cite{bd,golter1,bock3} show that only
a gauge field mass counterterm is needed. $S_{gf}$ is the gauge
fixing term which is not just a naive lattice transcription of the
continuum covariant gauge fixing, it has in addition appropriate
irrelevant terms. As a result, $S_{gf}$ has a unique absolute
minimum at $U_{\mu x}=1$, validating weak coupling perturbation
theory around gauge coupling $g=0$. For an explicit expression of
$S_{gf}$, see \cite{golter1}.

Obviously, the action $S_B(U)$ is not gauge invariant. By giving
it a gauge transformation the resulting action is
$S_B(\varphi^\dagger_x U_{\mu x} \varphi_{x+\hat{\mu}})$. By
restricting to the trivial orbit, $U_{\mu x}=1$, we arrive at the
reduced model action
\begin{equation}
S_{reduced} = S_F(U=1) + S_B(\varphi^\dagger_x \;1 \;\;
\varphi_{x+\hat{\mu}}) \label{reduced}
\end{equation}
where $S_F(U=1)$ is obtained quite easily from eq.(\ref{wgact}) and
\begin{equation}
S_B(\varphi^\dagger_x \;1 \;\;\varphi_{x+\hat{\mu}})
= -\kappa \sum_x \varphi^\dagger_x(\Box \varphi)_x + \tilde{\kappa}
\sum_x \left[\varphi^\dagger_x(\Box^2 \varphi)_x - B^2_x \right]
\label{redB}
\end{equation}
now is a higher-derivative scalar field theory action and $B_x$
is given by,
\begin{equation}
B_x = \sum_\mu \left( \frac{V_{\mu \,x-\hat{\mu}} + V_{\mu x}}{2}
\right)^2 \;\;\;\;\; {\rm with} \;\;\;\;\;\;  V_{\mu x} = {\rm Im} \;
\varphi^\dagger_x\varphi_{x+\hat{\mu}}.
\end{equation}
Perturbation theory around $g=0$ translates in the reduced model to
the same around $\tilde{\kappa}=\infty$.

In the following, we investigate the action (\ref{reduced}) at
small nonzero $y$ by numerical methods. To follow up the numerical
investigation, analytic studies will then be made again at small
$y$ with slightly different boundary conditions. 

\section*{Numerical results at small Yukawa coupling} \label{num}
At $y=0$, the domain wall and the anti-domain wall are detached
from each other. In fact, in this case there is no coupling between
the 4-dimensional worlds living on the $s_0$-th and the $(s_0+1)$-th
slices (and similarly for the $s_1$-th and the $(s_1+1)$-th slices).
From fermion current considerations, the LH chiral mode bound to
the domain wall at $s=0$ would then necessitate the generation of
a RH mode at $s=s_0$ and also at $s=s_1+1$. Similarly a LH mode,
mirror to the RH chiral mode at the anti-domain wall at $s=L_s/2$,
would be generated at $s=s_0+1$ and also at $s=s_1$. Our numerical
simulations at $y=0$ and similar studies in \cite{golter2} clearly
show that mirror chiral modes form at the waveguide boundaries.
Generation of these mirror modes at the waveguide boundaries is
independent of gauge fixing.

On the other hand, with gauge fixing at $y=1$ \cite{bd}, the mirror
chiral modes are certainly absent at the waveguide boundaries. In
this case, the only chiral modes are at the domain wall and at the
anti-domain wall and the spectrum is that of a free domain wall fermion.

To investigate the interesting question what happens at small $y$, 
we looked for chiral modes at $y=0.75,\,0.5,\,0.25$, especially at
the waveguide boundaries. Specifically we numerically evaluate the
$LL$ and the $RR$ propagators in momentum space conjugate to the 4
space-time dimensions and in configuration space for the 5-th
(flavor-like) dimension. This is done using quenched configurations
of the scalar fields at $\kappa = 0.05$ and $\tilde{\kappa}=0.2$.
This choice of the coupling parameters puts the theory quite close to
the phase transition from a ferromagnetic broken phase (FM) to a
rotationally noninvariant broken phase (FMD), staying within the
FM phase (in the full theory with dynamical gauge fields FM-FMD is
the transition where gauge symmetry is restored).
For a detailed discussion of the phase diagram of the quenched
reduced model, please see \cite{bock3}. The ($\kappa, \tilde{\kappa}$)
point we have chosen for the present investigation is also
where we had previously done all our work at $y=1$.

On a $6^3 \times 16$ lattice with $L_s=22$ and $m_0=0.5$, the
chiral propagators at the waveguide boundaries showed an increasing
trend as the fourth component of momentum $p_4$ (with $\vec{p}=
\vec{0}$) was decreased to the minimum value possible for this
lattice size, something that could signify a pole at zero momentum,
thereby indicating possible presence of mirror modes. The lower the
Yukawa coupling, the sharper was the increasing trend of the chiral
propagators with decreasing $p_4$. 

For a closer scrutiny we then took lattice sizes which were bigger in
the 4-th direction to accommodate lower $p_4$-values. We found that
for each $y$ there is a big enough lattice size for which the
chiral propagators ultimately start showing a decreasing trend,
indicating possible absence of mirror modes. Obviously the smaller 
the Yukawa coupling, the bigger the $L_4$ extension was required.
Moreover, all the chiral propagators at these small Yukawa couplings
matched exactly with the corresponding free chiral propagators. These
free chiral propagators were numerically determined by inverting
the free fermionic action (obtained by putting $\varphi =1$ in the 
reduced fermionic action $S_F(U=1)$) with the respective values of
the Yukawa couplings (the free fermionic action at $y=1$ is the
free domain wall action of Kaplan \cite{kaplan1}).

\begin{figure}
\vspace{-1.5cm}
\parbox{8cm}{\psfig{figure=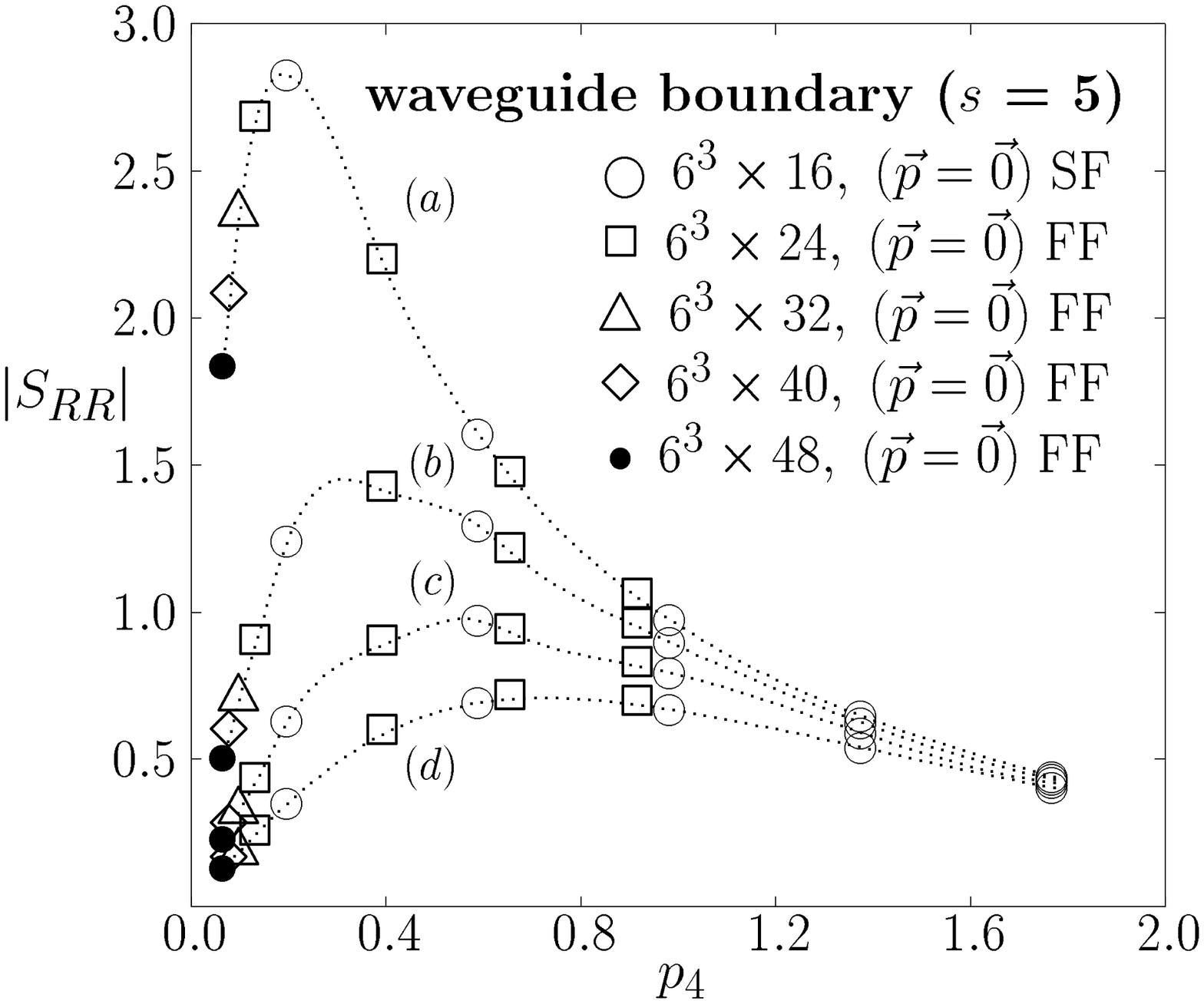,width=8.6cm,height=10.6cm}} \ \
\parbox{8cm}{\psfig{figure=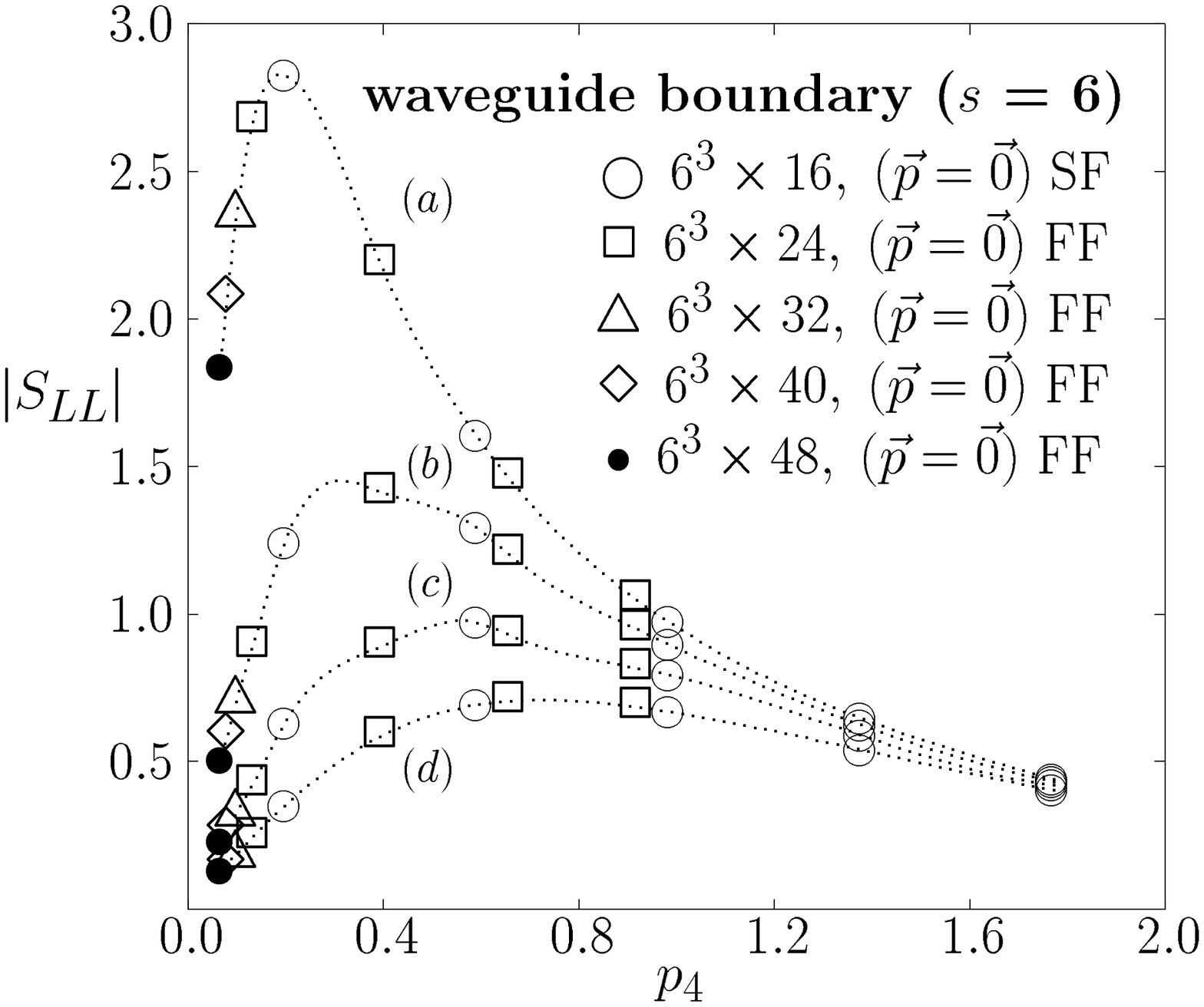,width=8.6cm,height=10.6cm}}
\vspace{-2.9cm}
\caption{$RR$ propagator at waveguide boundary $s=5$ and $LL$
propagator at waveguide boundary $s=6$ at $\kappa=0.05$ and 
$\tilde{\kappa}=0.2$ ($L_s=22$; a.p.b.c. in $L_4$),
(a) $y=0.25$, (b) $y=0.50$, (c) $y=0.75$, (d) $y=1.0$.}
\label{wgby}
\end{figure}
\vspace{0.0cm}
 
Our results at the waveguide boundaries $s = s_0 \equiv 5$, and
$s = s_0+1 \equiv 6$ are summarized
in Fig.\ref{wgby} for the $RR$ and the $LL$ propagators respectively.
In the figures, SF and FF respectively indicate data obtained by
numerical simulation of the scalar-fermion reduced model,
eq.(\ref{reduced}), and by direct numerical inversion of the free
fermion matrix (i.e. with $\varphi=1$) at the given Yukawa couplings.
Error bars are smaller than the symbols. Dotted lines joining the
data points are to guide the eye. Figure \ref{wgby} also contains
our previous data at $y=1$ \cite{bd} for comparison. From the figures,
existence of poles at zero 4-momentum for these chiral propagators
does not seem likely.

Similar numerical investigation is also carried out at the other
waveguide boundary $s_1$, $s_1+1$. There too evidence for mirror
modes is dim.

\section*{Perturbation Theory at $y \ne 1$} \label{bfdwa}
To confirm the indication from our numerical simulation that, for
small $y$, the chiral propagators at the waveguide boundaries do
not seem to have poles at zero momentum, it would be nice to
calculate the propagators analytically. This can be done in
perturbation theory in the coupling $1/\sqrt{2 \tilde{\kappa}}$.
This was also done for the fermion propagators to 1-loop in our
previous investigation at $y=1$ \cite{bd}. 

The present case, however, is more complicated due to the presence
of two more defects created at the two waveguide boundaries because
of $y\ne 1$, in addition to the existing defects, namely, the domain
wall at $s=0$ and the anti-domain wall at $s=L_s/2$. The defects are
the places across which translational invariance is not maintained
and boundary conditions need to be imposed to match the propagators
from two sides of each defect.   

In this section we leave Kaplan boundary conditions and use Shamir
boundary conditions \cite{shamir} because imposing boundary conditions
in the Shamir case would be a lot less tedious and our conclusions
would qualitatively be the same as in the Kaplan case.   

The free Dirac operator in the Shamir case is defined on 4 space-time
dimensions plus a finite 5-th dimension, $0 \leq s \leq L_s-1$ with
rigid walls at $s=0$ and $s=L_s-1$. The domain wall mass
is taken as $(-m_0)$ for all $s-$slices. Fermionic spectrum then
consists of a LH chiral mode at $s=0$ and a RH chiral mode at
$s=L_s/2$. (For further discussion see \cite{shamir} and \cite{aoki}.)

The waveguide in this case can be implemented by putting the
same 4-dimensional gauge field over $s$-slices from
$s_0+1$ to $L_s-1$ with a view to gauging only the RH mode at
$s=L_s-1$. The hopping terms from $s_0$ to $s_0+1$ again break the
local gauge invariance of the action. Longitudinal gauge degrees of
freedom or radially frozen scalar fields $\varphi$ enter the action
explicitly after a gauge transformation. The reduced model is then
obtained by imposing $U_{\mu x}=1$:
\begin{eqnarray}
S^{(shamir)}_{F}(U=1) & = & \sum_s \overline{\psi}^s \left(
\partial\!\!\!/ - w - m_0 \right) \psi^s + \sum_s \overline{\psi}^s
\psi^s - \sum_{s\ne s_0} \left( \overline{\psi}^s P_L \psi^{s+1} +
\overline{\psi}^{s+1} P_R \psi^s \right) \nonumber\\
&& - \; y \left( \overline{\psi}^{s_0} \varphi^\dagger P_L \psi^{s_0+1}
+ \overline{\psi}^{s_0+1} \varphi P_R \psi^{s_0} \right). \label{bfact}
\end{eqnarray}

The fermion propagators are obtained in momentum space for 4
space-time dimensions while staying in the coordinate space for
the 5-th dimension. The procedure in the following is applicable
to all values of the Yukawa coupling $y$ including $y=0$ and $y=1$.
Needless to say, it is also applicable to Kaplan boundary conditions.

In order to develop perturbation theory, in reduced model, we expand,
\begin{equation}
\varphi_x = \exp(ib\theta_x) = 1 +ib \theta_x  -\frac{1}{2} b^2
\theta_x^2 + {\cal O}(b^3), \;\;\;\;\;  b = \frac{1}{\sqrt{2
\tilde{\kappa}}}.
\end{equation}

Thus the fermion action at the tree level is written as,
\begin{equation}
S_F^{(0)} = \sum_{p,s,t} \overline{\widetilde{\psi_p^s}} \left[
i\overline{p\!\!\!/} \delta_{s,t} + \left( M_{\bar{y}} \right)_{st}P_L
+ \left( M^\dagger_{\bar{y}} \right)_{st} P_R \right]
\widetilde{\psi_p^t} \label{bfmom}
\end{equation}
where $\overline{p}_\mu = \sin(p_\mu)$ and $\overline{p\!\!\!/} =
\gamma_\mu\overline{p}_\mu$, and
\begin{eqnarray}
\left( M_{\bar{y}} \right)_{st} &=& M_{st} + \bar{y} \delta_{s,s_0}
\delta_{t,s_0+1} \nonumber\\
\left( M^\dagger_{\bar{y}} \right)_{st} &=& M^\dagger_{st} + \bar{y}
\delta_{s,s_0+1} \delta_{t,s_0} \label{My}
\end{eqnarray}
where $\bar{y} = 1 - y$. The $M$ and $M^\dagger$ are defined as,
\begin{eqnarray}
M_{st} &=& [ 1 - m_0 + \sum_\mu (1 - \cos p_\mu) ]
\delta_{s,t} - \delta_{s+1,t} \nonumber\\
M^\dagger_{st} &=& [ 1 - m_0 + \sum_\mu (1 - \cos p_\mu) ]
\delta_{s,t} - \delta_{s-1,t}. \label{Mst}
\end{eqnarray}

The free fermion propagator can now formally be written as,
\begin{eqnarray}
\Delta(p) &=& \left[ i\overline{p\!\!\!/} + M_{\bar{y}} P_L +
M^\dagger_{\bar{y}} P_R \right]^{-1} \nonumber \\
&=& \left( -i\overline{p\!\!\!/} + M^\dagger_{\bar{y}} \right) P_L
G_L(p) + \left( -i\overline{p\!\!\!/} + M_{\bar{y}} \right) P_R G_R(p)
\label{fprop}
\end{eqnarray}
where,
\begin{eqnarray}
G_L(p) &=& \frac{1}{\sum_\mu \overline{p}_\mu^2 + M_{\bar{y}}
M^\dagger_{\bar{y}}} \label{gl} \\
G_R(p) &=& \frac{1}{\sum_\mu \overline{p}_\mu^2 + M^\dagger_{\bar{y}}
M_{\bar{y}}}. \label{gr}
\end{eqnarray}
General solution of $G_L$ is obtained by writing (\ref{gl}) explicitly:
\begin{equation}
\sum_{s^\prime} \left[ \left( \overline{p}^2+ M M^\dagger
\right)_{s s^\prime} + \bar{y} \left( M_{s,s_0+1}\delta_{s^\prime,s_0}
+ \delta_{s,s_0} M^\dagger_{s_0+1,s^\prime} \right) + \bar{y}^2
\delta_{s,s_0} \delta_{s^\prime,s_0} \right] G_L(p)_{s^\prime,t}
= \delta_{s,t} \label{green}
\end{equation}
and similarly for $G_R$. We show only the calculations for obtaining
$G_L$ and henceforth drop the subscript $L$. 

We now define two translationally invariant regions $I$ and $II$
and set appropriate notation for $G$,
\begin{eqnarray}
{\rm region}\;\; I &:& \hspace{0.8cm} 0 \leq s \leq s_0, \hspace{2.4cm}
G = G^{(1)}, \\
{\rm region}\;\; II &:& \hspace{0.8cm} s_0+1 \leq s \leq L_s-1,
\hspace{1.0cm} G = G^{(2)}.
\end{eqnarray}
$G^{(1)}$ and $G^{(2)}$ are now forced to satisfy the translationally
invariant equation (with $B = 1- m_0 + \sum_\mu (1-\cos p_\mu)$),
\begin{equation}
\left[ \overline{p}^2 + 1 + B^2 \right] G^{(1,2)}_{s,t} -
B G^{(1,2)}_{s+1,t} - B G^{(1,2)}_{s-1,t} = \delta_{s,t}.
\label{greensp}
\end{equation}
The above equation (\ref{greensp}) actually takes the system beyond
the respective regions $I$ and $II$. This will be taken care of by
boundary conditions later.

The solutions of eq.(\ref{greensp}) are expressed as sum of
homogeneous and inhomogeneous solutions,
\begin{eqnarray}
G^{(1)}(p)_{s,t} &=& g^{(1)}(t)e^{-\alpha s} + h^{(1)}(t)
e^{\alpha s} + \frac{\cosh[\alpha (l - \vert s-t \vert )]}
{2B\sinh(\alpha)\sinh(\alpha l)}, \label{lmsol1} \\
G^{(2)}(p)_{s,t} &=& g^{(2)}(t)e^{-\alpha s} + h^{(2)}(t)
e^{\alpha s} + \frac{\cosh[\alpha (l - \vert s-t \vert )]}
{2B\sinh(\alpha)\sinh(\alpha l)}, \label{lmsol2}
\end{eqnarray}
where $l = L_s/2$, and
\begin{equation}
\cosh(\alpha)=\frac{1}{2}\left( B +\frac{1+\overline{p}^2}{B} \right).
\end{equation}
The third terms in eqs.(\ref{lmsol1}, \ref{lmsol2}) are the
inhomogeneous solutions, which are the same in both the regions.
Avoiding singularity in $\alpha$ when $B$ is zero restricts the
allowed range of $m_0$ to $0<m_0<1$. In this work we have taken
$m_0=0.5$.

The unknown functions $g^{(1,2)}(t)$ and $h^{(1,2)}(t)$
are determined from the following boundary conditions:
\begin{eqnarray}
G^{(1)}_{-1,t} &=& 0 \nonumber\\
G^{(2)}_{L_s-1,t} - B G^{(2)}_{L_s,t} &=& 0 \nonumber \\
\overline{{\cal F}}\, G^{(1)}_{s_0,t} -B G^{(1)}_{s_0-1,t}
- (1-\bar{y}) B G^{(2)}_{s_0+1} &=& \delta_{s_0,t} \nonumber\\
{\cal F} G^{(2)}_{s_0+1,t} -B G^{(2)}_{s_0+2,t} - (1-\bar{y})
B G^{(1)}_{s_0,t} &=& \delta_{s_0+1,t} \label{dwbc}
\end{eqnarray}
where $\overline{\cal F} = {\cal F}-\bar{y}(2-\bar{y})$ and
${\cal F} = \overline{p}^2 + 1 + B^2$. Substituting $2B\sinh(\alpha)
\sinh(\alpha l) = X^{-1}$ and using the boundary conditions,
eqs.(\ref{dwbc}), we arrive at,
\begin{equation}
{\bf A}\cdot{\bf v}(t) = {\bf X}(t), \label{mateq}
\end{equation}
where, ${\bf v}(t) = (g^{(1)}\;\;\; h^{(1)}\;\;\;
g^{(2)}\;\;\; h^{(2)})$ is a 4-component vector,
${\bf X}(t)$ is another 4-component vector and ${\bf A}$ is
a $4\times 4$ matrix as given below,

\[
{\bf A} = \left( \begin{array}{cccc}
e^{\alpha} & e^{-\alpha} & 0 & 0 \\
0 & 0 & e^{-\alpha(L_s-1)} - B e^{-\alpha L_s} &
e^{\alpha(L_s-1)}-B e^{\alpha L_s} \\
\overline{{\cal F}}\, e^{-\alpha s_0} - B e^{-\alpha (s_0-1)} &
\overline{{\cal F}}\, e^{\alpha s_0} - B e^{\alpha (s_0-1)} &
(\bar{y}-1) e^{-\alpha (s_0+1)} & (\bar{y}-1) e^{\alpha (s_0+1)} \\
(\bar{y}-1) e^{-\alpha s_0} & (\bar{y}-1) e^{\alpha s_0} &
{\cal F} e^{-\alpha (s_0+1)} - B e^{-\alpha (s_0+2)} &
{\cal F} e^{\alpha (s_0+1)} - B e^{\alpha (s_0+2)}
\end{array} \right)
\]

\noindent
and

\[
{\bf X}(t) = \left( \begin{array}{ll}
-X \cosh[\alpha(l- \vert -1-t \vert)] \hspace{2.0cm} & \\
& \\
B X \cosh[\alpha(l- \vert L_s-t \vert)] & \hspace{-2.2cm} -\;
X \cosh[\alpha(l- \vert L_s-1-t \vert)] \\
& \\
\delta_{s_0,t} + B X \cosh[\alpha(l- \vert s_0-1-t \vert)] &
\hspace{-0.6cm} -\; \overline{{\cal F}}\, X \cosh[\alpha(l- \vert
s_0-t \vert)] \;+ \\
& \hspace{2.0cm} (1-\bar{y}) X \cosh[\alpha(l- \vert s_0+1-t \vert)] \\
& \\
\delta_{s_0+1,t} + B X \cosh[\alpha(l- \vert s_0+2-t \vert)] &
\hspace{-0.3cm} -\; {\cal F} X \cosh[\alpha(l- \vert s_0+1-t \vert)] \;+ \\
& \hspace{2.0cm} (1-\bar{y}) B X\cosh[\alpha(l- \vert s_0-t \vert)]
\end{array} \right).
\]

The solution to the eqs.(\ref{mateq}) is very complicated in general,
particularly for finite $L_s$. However, $g^{(1,2)}(t)$ and
$h^{(1,2)}(t)$ can be obtained for finite $L_s$ by solving the above
equations numerically for different $t$ values. This way we can
easily construct the tree level fermion propagators at any given
$s$-slice.

The solutions for $(G_R)_{s,t}$ and the resulting propagators are
obtained in exactly the same way.

1-loop corrections to the tree level $LL$ and $RR$ propagators
above can be easily found following \cite{bd} and as in $y=1$
are found to be negligible. This is because the scalar fields are
almost completely decoupled.

\begin{figure}
\vspace{-1.2cm}
\hspace{4.0cm} \psfig{figure=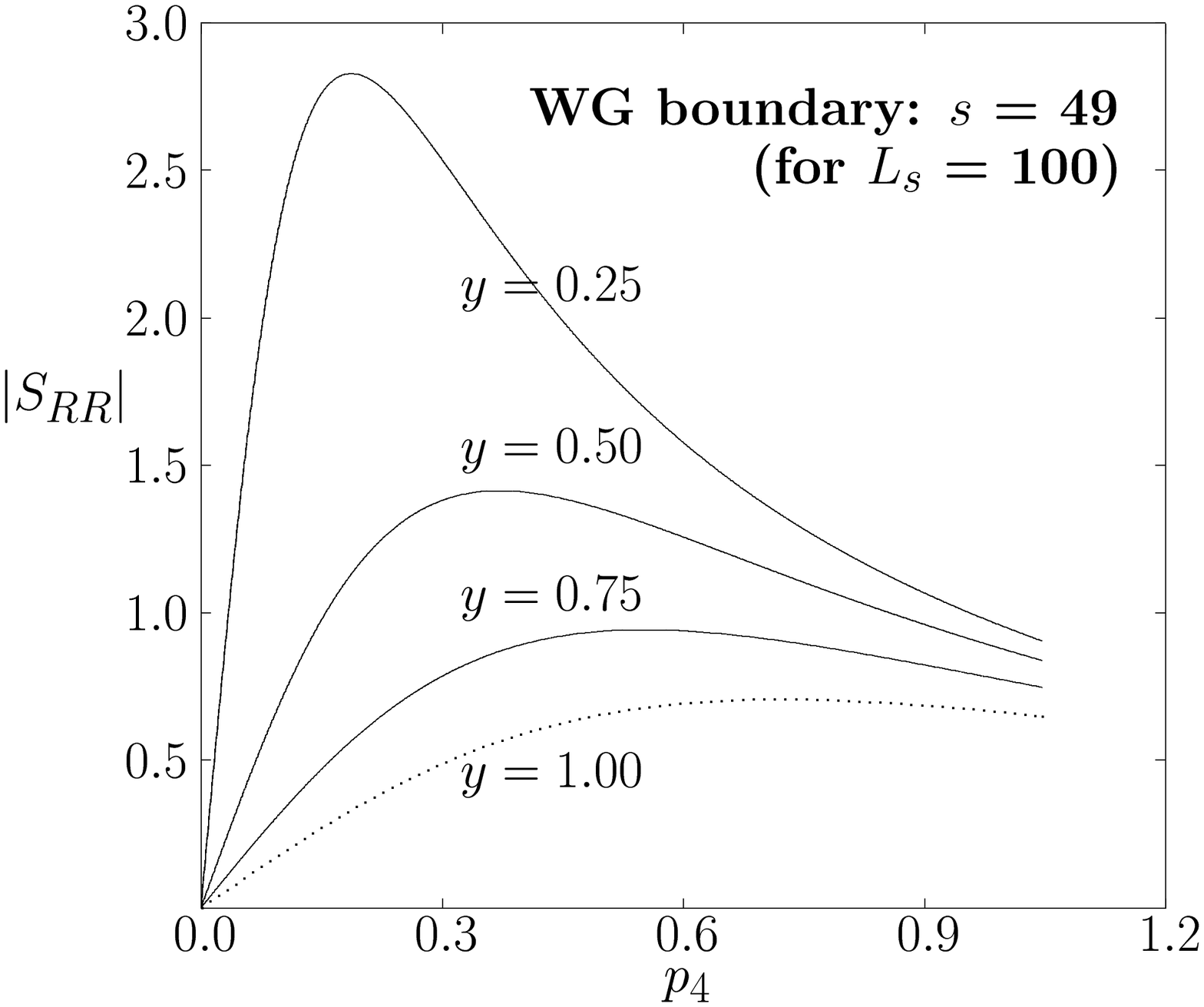,width=8.6cm,height=10.6cm}
\vspace{-2.9cm}
\caption{Tree level $RR$ propagator at waveguide boundary $s=49$
($512^3 \times 8192$ lattice, $L_s=100$; a.p.b.c. in $L_4$).}
\label{wgb}
\end{figure}

Using the analytic method described above, on a $512^3 \times 8192$
lattice with $L_s=100$ and $m_0=0.5$, we plot in Fig.\ref{wgb} tree
level $RR$ propagators for $y = 0.25,\,0.50,\,0.75,\,1.0$ at the
waveguide boundary $s = s_0 = 49$. From the figure it is clearly
seen that there are no poles in these propagators, ruling out any
possibility of a mirror mode at the waveguide boundary at any
nonzero $y$ (including arbitrarily small $y$).

\section*{Discussion} \label{dis}
In our previous study \cite{bd} we investigated the U(1) gauge
fixed domain wall waveguide model in the reduced limit at Yukawa
coupling $y=1$. Although the theory contained scalars explicitly
in the action as a result of gauge noninvariance, we found no
evidence of them. The fermionic spectrum was that of a free domain
wall fermion.

Since the Yukawa interaction couples only the two slices across a
waveguide boundary, there is a new defect introduced at a
waveguide boundary when $y\ne 1$. Especially at $y=0$ mirror chiral
modes are developed at the waveguide boundaries. It is an interesting
question to find out if these mirror modes persist to exist even
for small values of the Yukawa coupling. 

Our numerical investigations (at small $y$) of chiral propagators
at the waveguide boundaries initially show a disturbing trend of
increasing with decreasing momentum. On bigger lattices and smaller
momenta, however, this trend does not continue and poles of the
chiral propagators do not seem likely. In addition, the data agree
quite well with free chiral propagators obtained by numerically
inverting the fermion matrix after putting $\varphi=1$  for all
the $y$ values. Agreement with free propagators means that the
scalar fields are decoupled.

To be able to go to much smaller momentum, we then perform an
analytic calculation of the propagators at $y \ne 1$ in the
perturbative limit $\tilde{\kappa}=\infty$. Although done with
different boundary conditions than we started with, qualitative
conclusions can at least be made and the analytic tree propagators
certainly show a very good resemblance to the numerically calculated
ones and confirm the view that there are no poles for these
propagators at the waveguide boundaries.

Our conclusion would be that even for arbitrarily small or any
nonzero Yukawa coupling, the fermion spectrum
would be devoid of any ill-effects of the scalar fields. In fact the
scalar fields are completely decoupled. This happens because the
continuum limit is taken at the FM-FMD transition where the scalars
do not scale, being far away from a ferromagnetic-paramagnetic transition.
Obviously from a quenched numerical simulation alone, we cannot
come to a firm conclusion. However, the 1-loop corrections to the
perturbative tree level fermion propagators are negligibly small and
fermion loops enter the calculation at least at the two loop level.
Hence our conclusions about nonexistence of the mirror modes at
the waveguide boundaries for small $y$ seem to be correct even
for the situation with dynamical fermions as long as the phase diagram
stays qualitatively the same with dynamical fermions. Once the
continuum limit can be taken at  the FM-FMD transition, even an
arbitrrily small $y$ (which effectively mimics a scaling radial mode
of the scalar fields) cannot produce the mirror modes and spoil the theory.



\begin{thebibliography}{99}

\bibitem{bock1} W. Bock, M.F.L. Golterman, Y. Shamir, Phys.
Rev. Lett 80 (1998) 3444
\bibitem{bd} S. Basak, A.K. De, Phys. Rev. D64 (2001) 014504
\bibitem{golter1} Y. Shamir, Phys. Rev. D57 (1998) 132; M.F.L.
Golterman, Y. Shamir, Phys. Lett. B399 (1997) 148
\bibitem{kaplan1} D.B. Kaplan, Phys. Lett. B288 (1992) 342
\bibitem{kaplan2} D.B. Kaplan, Nucl. Phys.B (Proc. Suppl.)
30 (1993) 597; M.F.L. Golterman, Y. Shamir, Phys. Rev. D51
(1995) 3026
\bibitem{golter2} M.F.L. Golterman, K. Jansen, D.N. Petcher, J.C.
Vink, Phys. Rev. D49 (1994) 1606
\bibitem{bock3} W. Bock, M.F.L. Golterman, Y. Shamir, Phys. Rev.
D58 (1998) 54506
\bibitem{shamir} Y. Shamir, Nucl. Phys. B406 (1993) 90
\bibitem{aoki} S. Aoki, Y. Taniguchi, Phys. Rev. D59 (1999) 54510
\end{thebibliography}
\end{document}